\documentstyle[sprocl]{article}
\def\ra{\rightarrow}

\def\be{\begin{equation}}
\def\ee{\end{equation}}
\def\bea{\begin{eqnarray}}
\def\eea{\end{eqnarray}}

\begin{document}
\begin{flushright}
UH-511-885-98   \\
January 1998
\end{flushright}
\vspace{.25in}
\title{NEUTRINOS\footnote{Based on lectures delivered at the ICTP
Summer School on High Energy Physics and Cosmology, Trieste, June 16-20, 1997.}}
\author{S. PAKVASA}
\address{Department of Physics \& Astronomy,\\
University of Hawaii \\
Honolulu, HI  96822 USA}
%
%

\maketitle


\section{Introduction}

We can start with the question:  why are neutrino properties especially
interesting?  Recall that in the minimal standard model there are no
right handed neutrinos and furthermore lepton number is
conserved so neutrinos can have neither Dirac nor Majorana masses.
Furthermore, with zero masses, the mixing angles in the charged weak
current are all zero.  Any evidence for non-zero masses or mixing
angles is  evidence for physics beyond the standard model and hence
potentially a powerful tool.  Besides, the masses and mixing angles are
fundamental parameters which will have to be explained by the eventual
theory of fermion masses. Massless neutrino can be guaranteed by
imposing chiral symmetry.
Since chiral symmetry and just masslessness are difficult to distinguish
at present, one can ask if there are any other fundamental reasons for
$m_\nu$ to be zero, such as gauge invariance for the photon and graviton
masses.  The
only other massless particles we know are Nambu-Goldstone particles due
to spontaneous breaking of global symmetries.  If $\nu$ was such a
particle it would obey soft-$\nu$ theorems analogues of soft $\pi$
theorems \cite{wit}.
Hence any amplitude A with $\nu$ as an external particle of momentum $q$
should satisfy
\begin{eqnarray}
\begin{array}{l}
1 im  \ A (\nu) = 0  \\ \nonumber
q_\mu \rightarrow 0 
\end{array}
\end{eqnarray} 
i.e. A should vanish linearly as q.  This additional dependence of
amplitudes on the neutrino momentum, for example, would distort Kurie
plots for beta decay grossly from linearity.  Since no such deviations
are observed we can rest assured that neutrino is not a Nambu-Goldstone
particle.  Hence the neutrino is neither a gauge particle nor a
Nambu-Goldstone particle and is not required to be massless.

Once neutrinos have masses, by analogy we expect that the only
distinguishing feature between flavors is different masses and mixings.
Hence we would expect $\nu_e, \nu_\mu$ and $\nu_\tau$ (or the
corresponding mass eigenstates $\nu_1, \nu_2$ and $ \nu_3)$ to have
different masses in general.

Let me review briefly the kinds of neutrino masses that can arise.  Let
$\psi_L$ and $\phi_R$ be two-component chiral (Weyl) fields; with
$\chi_R \equiv \psi^c_L$ and $\xi_L \equiv \phi^c_R$ the
charge conjugates.  Then there are two kinds of Majorana mass terms
possible:
\begin{eqnarray}
m_L \ \bar{\psi}_L \ \chi_R  + \ h.c. \quad \quad  \quad \quad (a)\\ \nonumber
m_R \ \bar{\phi}_R \ \xi_L    + \ h.c. \quad \quad \quad \quad (b)
\end{eqnarray}
These violate lepton number by 2 units and the term (2a) violates weak
SU (2) and has $\Delta I = 1$ whereas (2b) has $\Delta I = 0.$  
The Dirac mass term transforming as $\Delta I = \frac{1}{2}$ is given by
\begin{equation}
m_D \left [ \bar{\psi}_L \ \phi_R \  \  + h.c.  \right ]
\end{equation}
and conserves lepton number.

	In the standard model, for example, with particle assignments:

\begin{equation}
\left (
\begin{array}{c}
\nu_e \\ e 
\end{array}
\right )_L \ e_R   
\end{equation}
the electron gets a Dirac mass $m_e (\bar{e}_L e_R + h.c.)$.  
If there is a $\nu_{eR}$
and $\nu_e$ also gets a Dirac mass $m_{\nu_{e}} \left ( \bar{\nu}_{e_{L}}
\nu_{e_{R}} + h.c. \right )$ then it is difficult to understand why
$m_{\nu_{e}}/m_e$ is such a small number $(< 10^{-5}).$
At least so runs an oft-quoted sentiment  (One should bear in 
mind that $m_e/m_t < 10^{-5}$ and is equally hard to understand!)  
In any case, $m_i$ for all neutrinos are very small compared to 
other fermion masses.  How is it possible to get very small 
masses for neutrinos?   Some of the possibilities that
have been discussed in the literature are:

i)  Arrange the theory so that at tree level \underline{and} at one loop 
level $m_\nu = 0$.  The first non-zero contribution arises at 
two-loop level and hence one expects $m_{\nu e} \approx \alpha^2 m_e$ 
which gives $\nu_e$ mass in eV range!  
Similarly $m_{\nu \mu}$ and $m_{\nu \tau}$ would be in 100 eV 
and few KeV range.  Whether the neutrinos are Dirac or 
Majorana particles depends on the detailed model.

ii)  The See-Saw Mechanism \cite{yanagida}:  In the general case when both Dirac 
and Majorana mass terms are present, 
there is a 2x2 mass matrix for every flavor.
\begin{equation}
\left ( \bar{\psi}_L \ \bar{\xi}_L \right ) 
\left [
\begin{array}{cc}
 m_L & m_D \\
m_D & m_R 
\end{array} \right ] 
\left [
\begin{array}{c}
\chi_R \\
  \phi_R 
\end{array} \right ]
\end{equation}

This has eigenvalues $m_1$ and $m_2$ and the eigenstates 
$\chi_1, \chi_2$ are Majorana.  The case when $m_1 = m_2$ 
is the special one which allows for a conserved lepton number 
to be defined and gives a 4 component Dirac particle.  In the 
case where $m_R$ is much larger than the others, e.g. 
$m_R >> m_D >> m_L$ the e. values simplify to
\begin{eqnarray}
m_1 & \cong & m_D^2/m_R  \\ \nonumber
m_2 & \cong & m_R
\end{eqnarray}
and $m_1 << m_D$.  So $m_\nu$ can be much smaller than a typical 
$m_D$ say $m_e$.

In this case,
i)  $\nu's$ are Majorana particles and
ii)  the mass hierarchy in $\nu's$ is expected to be
\begin{equation}
m_{\nu_{e}} / m_{\nu_{\mu}} / m_{\nu_{\tau}} \sim 
m^2_e / m^2_\mu  / m^2_\tau
\end{equation}
or $m^2_u/m^2_c/m^2_t$ etc.
 
Let me remind you of some specific scenarios for neutrino masses:  
(i) No $\nu_R's$ exist.  Lepton number is a global symmetry and 
spontaneously broken via the vacuum expectation value (vev) of 
an I=1 complex scalar field \cite{gelmini}.  The coupling of the neutral 
field so $\nu's$ give the left-handed $\nu's$ a Majorana mass 
$(m_L)$.  The massless Nambu-Goldstone field is the Majoron.  
The masses $m_\nu$ remain arbitrary.  One can also consider 
the I=1 complex field as a way of expressing a product of 
two $I= \frac{1}{2}$ ``standard" Higgs multiplets.  (b) 
$\nu_R$'s exist \cite{chicasige}.  This is the case in some unification 
groups, (SO (10), $E_6$ etc).  In one class 
of models $m_L \cong 0$ and $m_R \neq 0$ and corresponds 
to the usual see-saw mechanism.  In a more general case 
one can envisage $m_L \neq 0, m_R \neq 0$ and the 
possibility that $m_1$ and $m_2$ may be close and the
 mixing angle $\theta$ may be large!  Then for three 
flavors there is a 6x6 mixing matrix and mixing with 
3 sterile (I=0) neutrinos.

The current experimental limits on neutrino masses are \cite{lobashov}:
\begin{eqnarray}
m_{\nu_{e}} & < & 4 \ eV  \\ \nonumber
m_{\nu_{\mu}} & < & 170 \ KeV  \\ \nonumber
m_{\nu_{\tau}} & < & 18.2 \ MeV  
\end{eqnarray}
All of these are direct, kinematic limits from Laboratory experiments.  

With non-zero masses, in general there is mixing i.e. mass 
eigenstates may not be identical to weak eigenstates.
In the most general case the mixing can be between the 
three flavors $\nu_e, \nu_\mu \nu_\tau.$
For simplicity, let us first consider two flavor mixing between $\nu_e$ and
$\nu_\mu$:  
with the mass eigenstates $\nu_1$ and $\nu_2 (m_2 > m_1)$.   For 
two flavor mixing (say $\nu_e$ and $\nu_\mu)$.  The standard 
forms for survival probability and conversion probability are given by
\begin{eqnarray}
P_{ee} (L) & = & 1 - \sin^2 2 \theta \sin^2 \left (\frac{\delta m^2
L}{4E} \right ) \nonumber \\
P_{e \mu} (L) & = & \sin^2 2 \theta \sin^2 \left (\frac{\delta m^2
L}{4E} \right )
\end{eqnarray}
for a neutrino starting out as $\nu_e$.  Here $\theta$ 
is the mixing angle 
$\delta m^2 = m^2_2 - m^2_1, \ \ L =ct$ and the ultra-relativistic limit
$E = p+ \frac{m^2}{2p}$  has been taken.  Although these formulae \cite{kayser} 
are usually derived in plane wave approximation with $p_1 =p_2$, it has 
been shown that a careful wave packet treatment yields the same
results.  
When  the argument of the oscillating term $(\frac{\delta m^2 L}{4E})$ 
is too small, no oscillations can be observed.  When it is much larger 
than one then due to the spread of $E$ at the source or finite
 energy resolution of the detector the oscillating term 
effectively averages out to 1/2.

There are some obvious conditions to be met for oscillations to take
place \cite{kim}. As the beam travels, the wave packet spreads and the mass 
eigenstates separate.  If the width $\Delta x$ remains greater 
than the separation then oscillations will occur; but if the separation 
is greater, then two separate pulses of $\nu_1$ (mass $m_1)$ and $\nu_2$ 
(mass $m_2)$ register in the detector with intensities $\cos^2 \theta$ and
$\sin^2 \theta$ separated by $\Delta t = \frac{\delta m^2}{2 E^2} \
(L/c)$.  In principle, the 
intensities as well as oscillation expressions should reflect the
slightly 
different decay widths for different mass eigenstates but this is 
of no practical importance.  The same expressions remain valid if the 
mixing is with a sterile neutrino with no weak interactions.  
Since the sterile neutrino has no weak interactions, not even neutral 
current, there is an apparent non-conservation of probability.  
In general, to describe flavor mixing among three flavors, a 
3x3 analog of Kobayashi-Maskawa matrix is necessary.  To describe the 
general mixing of 3 flavors and 3 sterile states a 6x6 unitary 
matrix is called for.  In this case, the 3x3 flavor mixing matrix is, 
of course, not unitary due to leakage to the sterile sector.

In general, there will be CP violation due 
to phases in the mixing matrix \cite{cabibbo}.  The deviation of
$P(\nu_\alpha \ra \nu_\beta, t) -P (\bar{\nu}_\alpha \ra \bar{\nu}_\beta,t)$ 
from 0 is a measure of CP violation. $P(\nu_\alpha \ra \nu_\beta, t)$ and
$P (\bar{\nu}_\beta \ra \bar{\nu}_\alpha,t)$ are equal by CPT.  Another
way to check CP is follow a single probability over a long time
(distance) and do a Fourier analysis.  If it contains A+B  coswt + C
sinwt.. then $C \neq$  0 implies CP violation.  If $\Sigma_\alpha P \ (\nu_\alpha \ra
\nu_{\alpha } t) \leq \frac{4}{3}$ then either CP is violated or there
are more than three states mixing.  Another interesting test can be done
using $\nu, \bar{\nu}$ beams from bean dump or $K_L$ decay.  If $(\nu_e -
\bar{\nu}_e)/(\nu_e + \bar{\nu}_e)$ or $(\nu_\mu
-\bar{\nu}_\mu)/(\nu_\mu + \bar{\nu}_\mu)$ are
non-zero, CP is violated.  There are also T-violating corrections
in flavor changing decays such as $\mu \ra \bar{e}ee, \tau \ra \mu e
\bar{e}$ etc., of the kind $\sigma.  (P_1 \times P_2)$.
An old observation \cite{lee} which
has become relevant recently is the following:  it is possible for
neutrinos to be massless and not orthogonal.  For example, with three
neutrino mixing we have
\begin{eqnarray}
\nu_e & = & U_{e1} \nu_1  \ + U_{e2} \nu_2 \ + U_{e3} \nu_3  \\
\nonumber
\nu_\mu & = & U_{\mu 1} \nu_1  \ + U_{\mu 2} \nu_2 \ + U_{ \mu 3} \nu_3
\end{eqnarray}
Suppose $m_1 = m_2 = 0$ but $m_3$ is non-zero and $m_3 > Q$ where Q is
the energy released in $\beta-$decay or $\pi-$decay producing $\nu_e$ and $\nu_\mu$ beams.
Then $\nu_e$ and $\nu_\mu$ will have zero masses but will not be
orthogonal:
\begin{equation}
< \nu_e \mid \nu_\mu > =  - U^*_{e3} U_{\mu 3} \neq 0 
\end{equation}
(Scenarios similar to this are realized in combined fits to solar and
LSND neutrino anomalies).  Incidentally, the ``$\nu_e$'' and
$''\nu_\mu$'' produced in Z decay will not be massless and will be
nearly orthogonal!  This example illustrates the fact that neutrino
flavor is not a precise concept and is process dependent.

If a heavier neutrino $\nu_H$ is mixed with $\nu_e$ and $\nu_\mu$ with
mass in the range KeV to MeV it can show up in the abrupt changes of
phase space as the Q-value in a decay process passes $m_{\nu H}$.  This
simple and obvious idea was first exploited in 1963 \cite{nakagawa}.  
Typically one
expects kinks in energy spectrum in $\beta$-decay, $Ke_3, \pi_{e3} \ K
\mu_3$ etc.  So far such searches have yielded null results.  Similar
searches can also be made in 2-body decays such as $\pi \ra \mu \nu_\mu$ for
$\mu's$ of anomalous momenta.  

Oscillations of massless neutrinos arise under two circumstances.  One
is when gravitational couplings of neutrinos are flavor non-diagonal.
For example, $\nu_1$ and $\nu_2$ may couple to gravity with different
strengths:
\begin{equation}
H_{gr} = f_1 GE, \phi + f_2 \ G E_2 \phi
\end{equation}
where $\phi$ is the gravitational potential.  Then if $\nu_e$ and
$\nu_\mu$ are mixtures of $\nu_1$ and $\nu_2$ with a mixing angle $\theta$,
oscillations will occur with a flavor survival probability given by\cite{gasperini}
\begin{equation}
P = 1 - \sin^2 2 \theta \sin^2 \left (\frac{1}{2} \delta f \phi E L
\right )
\end{equation}
for a constant potential $\phi$.  If Lorentz invariance is violated with
neutrinos being velocity eigenstates corresponding to different maximum
velocities; then also there are oscillations with $P$ given by\cite{coleman}
\begin{equation}
P= 1- \sin^2 2 \phi  \ \sin^2 \left (\frac{1}{2} \  v EL \right )
\end{equation}
In both these cases, the dependence of oscillations is on (EL) to be
contrasted to L/E dependence of conventional oscillations.

{\underline{Matter Effects:}}  In traversing thru matter, the coherent
forward scattering of neutrinos with matter gives rise to effective
interaction energy which distinguishes $\nu_e$ from other flavors. The result is
\begin{eqnarray}
\delta H & = & \pm 
 \frac{\sqrt{2}G_F}{2m_N} \rho (Y_e - 1/2 Y_n) \nonumber \\
\mbox{for} \ \nu_e\mbox{'s and}  \\  
\delta H & = & \pm 
 \frac{\sqrt{2}G_F}{2m_N} \rho ( - 1/2 Y_n) \nonumber
\end{eqnarray}
for $\nu_\mu$'s and $\nu_\tau$'s.  The + sign for $\nu$'s and -sign for
$\bar{\nu}$'s, $\rho$ is the density of matter and $Y_e$ and $Y_n$
are the number of electron and neutrons respectively per nucleon.  There
is no such term for sterile neutrinos \cite{wolfenstein}.

As a result of this, the mixing angles and $\delta m^2$'s in matter are
different from their vacuum value.  The most interesting result is the
fact that for a given value of $\delta m^2$, mixing angle and
neutrino energy, there is always {\underline{some}} value of density 
$\rho$ where the matter angle becomes $45^0$ i.e. is maximal \cite{mikheyev}. This is
true no matter how small the vacuum mixing angle is.  Furthermore, this
enhancement can only occur for either $\nu$'s or for $\bar{\nu}$'s but
not for both.  Another related result is that (as long as $Y_e -
\frac{1}{2} Y_n > 0) \nu_e$ in matter has higher energy than
$\bar{\nu}_e$  and $\bar{\nu}_\mu$ has higher energy than $\nu_\mu$.
Hence if there are lepton-number violating couplings to say, a massless
Majoron, then decays such as\cite{berezhiani}
\begin{equation}
\nu_e \ra \bar{\nu}_e + M
\end{equation}
can occur in matter but not in vacuum.
Matter effects are important for solar neutrinos (if $\delta m^2 \sim
10^{-4}$ to $10^{-7} ev^2),$ for upcoming atmospheric neutrinos and can
be for supernova neutrinos.

$\nu-$ Decays:  For neutrinos below 1 MeV, the only decay modes possible
are 
\begin{eqnarray}
\begin{array}{l}
(1) \ \nu_\alpha \ra \nu_\beta + \gamma  \\  
(2) \ \nu_\alpha \ra \nu_\beta + \nu_i \bar{\nu}_i  \\ 
(3) \ \nu_\alpha \ra \bar{\nu}_\beta + M.  
\end{array}
\end{eqnarray}
The decay rate for the radiative mode, at one loop level, in the
standard model is given by
\begin{equation}
\Gamma_1 = \frac{G^2_F \ m_\alpha^5}{128 \pi^4}  
\left ( \frac{9}{16} \right ) \frac{\alpha}{\pi}
\left | \sum_{i} 
\left ( \frac{m_i}{m_W} \right )^2   
U_{i \alpha} \ U^*_{i \beta}  \right |^2
\end{equation} 
assuming $m_\alpha >> m_\beta$.  Cosmological and SN1987A limits on this
mode are of order $\tau/BR \ 10^{24}$s.  The rate for the $3 \nu$ mode
depends on whether it is one loop induced or whether there is GIM
violation at tree level
\begin{eqnarray}
\Gamma_2 (1 - loop) & = & \frac{G^2_F \ m_\alpha^5}{128 \pi^3}
\left ( \frac{\alpha^2}{8 \pi^2} \right ) \left |
\frac{\Sigma m_j^2}{m_W^2}
U_{j \alpha} \ U^*_{\beta j} \right |^2 \nonumber \\
\Gamma_2 (tree) & = & \frac{\epsilon^2 GF^2}{192 \pi ^3} \ m_\alpha^5
\end{eqnarray}
It is not easy to arrange for this decay rate to be significant.  The
decay rate for Majoron or familon decay mode depends on the unknown
Yukawa coupling, g,

Magnetic Dipole Moment:  A Majorana particle cannot have a non-zero magnetic dipole moment.
Hence for a non-zero magnetic dipole moment either neutrino is a Dirac
particle or the dipole moment is a transition moment between two
different Majorana states (e.g. $\nu_{eL}$ and $\nu^c_{{\mu}_{R}}).$  

In the standard model, the one loop calculation yields\cite{fujikawa}

\begin{equation}
\mu_{\nu_{e}} = \frac{3 m_e G_F}{4 \sqrt{2} \ \pi^2} \ m_{\nu_{e}} \mu_B
\end{equation}
which is $3.10^{-19} (m_\nu/eV) \mu_B.$  With mixing to a heavy lepton of
$m_L \sim 100$ GeV and mixing $\nu_e - \nu_L$ of 0.1, this can be
enhanced to $10^{-14} \mu_B.$  The simplest modifications of standard
of model which can yield large magnetic dipole moment for $\nu_e$ are
ones with extra scalar fields.

Double Beta Decay:  Of particular interest for neutrino properties is the neutrino-less
variety:
\begin{equation}
(A, Z) \ra (A, Z + 2) + e^- + e^-.
\end{equation}
This can only happen if $\nu^c \equiv \nu$ and $m_\nu \neq 0 ($ or
$\nu_e$ mixes with a massive Majorana particle).  The decay rate depends
on the $m_\nu$ and the nuclear matrix element \cite{haxton}.
\begin{equation}
\Gamma \cong \frac{1}{2.10^{15} yr}
\left [ M_{GT} \right ]^2   F_\nu   G_c
\end{equation}
where $G_c$ is the Coulumb correction factor, $F_\nu = m_\nu^2/m_e^2 \ f
(\epsilon_e/m_e)$ and $M_{GT}$ is the nuclear matrix element.
\begin{equation}
M_{GT} \sim < f \left | m \sum_{1} n \ \tau^+_m \ \tau^+_n \ \sigma_m
. \ \sigma_n
r^{-1}_{mn} \right | i >
\end{equation}    
This is in the limit of small $m_\nu$.  The strongest limits on are for Ge
of almost $10^{-24}$ yr.$^{-1}$.  Using calculated nuclear matrix
elements this places a limit on a Majorana mass for $\nu_e$ of $m_{\nu
e}^{M} \tilde{<} 1 eV$.  Eventually the matrix element can be extracted
from $2 \nu$ double $\beta$-decay.  With the next generation 
it may be possible to lower the bound to 0.2 - 0.3 eV.

\section{Atmospheric Neutrinos}

The cosmic ray primaries produce pions which on decays produce 
$\nu_\mu's$ and $\nu_e's$ by the chain $\pi \ra \mu \nu_\mu$,
$\mu \ra e \nu_e \nu_\mu.$  Hence, one expects a $\nu_\mu/\nu_e$ ratio of
2:1.  As energies increase the $\mu's$ do not have enough time (decay
length becomes greater than 15-20 km) and the $\nu_\mu/\nu_e$ ratio
increases.  Also at low energies the $\nu$ flux is almost independent
of zenith angle; at high energies due to competition between $\pi$-decay
and $\pi$-interaction the famous ``sec $(\theta)$'' effect takes over.
Since the
absolute flux predictions are beset with uncertainties of about 20\%, it is
better to compare predictions of the ratio (which may have only a 5\%
uncertainty)
$\nu_\mu/\nu_e$ to data in the form of  the famous double ratio
$R= (\nu_\mu/\nu_e)_{data} / (\nu_\mu /\nu_e)_{mc}$.

For the so-called ``contained'' events which for Kamiokande and IMB
correspond to visible energies below about 1.5 GeV, the weighted world
average (before SuperKamiokande) is $R = 0.64 \pm 0.06$ \cite{nakahata}.  This includes
all the data from IMB, Kamiokande, Frejus, Nusex and Soudan.  
The new SuperK results are completely
consistent with this \cite{totsuka}.  It may be worthwhile to recall all the doubts and
concerns which have been raised about this anomaly (i.e. deviation of R
from 1) in the past and their resolution.  (i) Since initially the
anomaly was only seen in
Water Cerenkov detectors, the question was raised whether the anomaly was
specific to water Cerenkov detectors.  Since then, it has been seen in a
tracking detector i.e. SOUDAN II. (ii) Related to the above was the
concern whether $e/\mu$ identification and separation was really as
good as claimed by Kamiokande and IMB.  The beam tests at KEK
established that this was  not a problem \cite{kasuga}.  (iii) The $\nu_e$ and
$\nu_\mu$ cross-sections at low energies are not well known; however
$e - \mu$ universality should hold apart from known kinematic effects. 
(iv) If more $\pi^{+'}s$ than $\pi^{-'}s$ are produced, then even though the
ratio of 2/1 is preserved there is an asymmetry in $\bar{\nu}_e/\nu_e$
versus $\bar{\nu}_\mu/\nu_\mu$.  Since $\nu$ cross-sections are larger than $\bar{\nu}$
cross-sections, the double ratio R would become smaller than 1 \cite{volkova}.
However, to explain the observed R, $\pi^{+'}s$ would have to dominate over
$\pi^{-'}s$ by 10 to 1, which is extremely unlikely and there is no
evidence for such an effect.  (v) Cosmic ray muons passing thru near (but
outside) the detector could create neutrals (especially neutrons) which
enter the tank unobserved and then create $\pi^{0'}s$ faking ``e'' like
events \cite{ryazhkaya}.  Again this effect reduces R.  However, Kamiokande plotted their
events versus distance from wall and did not find any evidence for more
``e'' events near the walls \cite{fukuda}. (vi) Finally, the measurement of $\mu$
flux at heights of 10-15 km to tag the parent particles as suggested by
Perkins was performed by the MASS collaboration \cite{bellotti}.  This should help decrease
the uncertainty in the expected $(\nu_\mu/\nu_e)$ flux ratio even
further.  It seems that the anomaly is real and does not have any
mundane explanation.  

We now turn to an explanation in terms of neutrino oscillations \cite{learned}.
Deviation of $R_{obs}/R_{MC}$ from 1 is fairly uniform over zenith angle
and is most pronounced in the charged lepton energy range 200-700 MeV
which corresponds to neutrino energies from 300 MeV t0 1.2 GeV.  If we
are to interpret this deficit of $\nu_e$'s (and/or excess of $\nu_e$'s)
as being due to neutrino oscillations, the relevant parameters are
determined rather easily.  The typical height of production, $h$, is
about 15-20 km above ground and for a zenith angle $\theta$ the distance
traveled by the neutrino before reaching the detector is
\begin{equation}
L(\theta) = R \left [ \sqrt{(1 + h/ R)^2 - \sin^2 \theta} - \cos \theta
\right ]
\end{equation}
where $R$ is the radius of the earth.  Allowing for angular smearing due
to the scattering and finite angular resolution one finds that neutrino
path lengths vary between 30 and 6500 km, and hence $L/E$ can vary
between 25 km/GeV and 20,000 km/GeV.  Since the data (pre-SuperK) 
did not show any $L$
(i.e. $\theta)$ or $E$ dependence one was led to infer that the oscillations
had already set in at $E_\nu \sim 1$ GeV and $L \sim 30 km$ and hence
$\delta m^2$ could not be much smaller than $10^{-2} eV^2$.  As for the
mixing angle $\theta$, if $P$ denotes the average oscillation
probability i.e. $P = sin^2 2 \theta < \sin^2 \delta m^2 L/4E > \approx
\frac{1}{2} \sin^2 2 \theta$; then $R=1  -P$ in case of $\nu_\mu -
\nu_\tau$ oscillations and for $\nu_\mu - \nu_e$ oscillations
\begin{equation}
R = \frac{1 - (1-r) P}{1 + (1/r -1) P}
\end{equation}
where $r = N (\nu_e)/N (\nu_\mu)$ in absence of oscillations and most
flux calculations yield $ r \sim$ 0.45.  Since $R$ is nearly 0.6, large
mixing angles of order $30^0$ to $45^0$ are called for, $\nu_\mu-\nu_e$ mixing
needing somewhat smaller one.  Detailed fits by Kamiokande and IMB, bear these expectations out although somewhat bigger range
of parameters $\delta m^2$  up to $4.10^{-3} eV^2$ and mixing angles up to
$20^0)$ are allowed.

	If the atmospheric neutrino anomaly is indeed due to neutrino
oscillations as seems more and more likely; one would like to establish
just what the nature of oscillations is.  There have been several
proposals recently.  One is to define an up-down asymmetry for $\mu's$
as well as $e's$ as follows \cite{flanagan}:

\be
A_\alpha = (N_\alpha^d - N_\alpha^u) / (N_\alpha^d + N_\alpha^u)
\ee
where $\alpha = e$ or $\mu$, $d$ and $u$ stand for downcoming
$(\theta_Z= 0$ 
to $\pi/2)$ and upcoming $(\theta_Z = \pi/2$ to $\pi)$ respectively.
$A_\alpha$ is a function of $E_\nu$.  The
comparison of $A_\alpha (E_\nu)$ to data can distinguish various
scenarios for $\nu$-oscillations rather easily \cite{flanagan}.  This asymmetry has the advantage
that absolute flux cancels out and that statistics can be large.  It can
be calculated numerically or analytically with some simple assumptions.
One can plot $A_e$ versus $A_\mu$ for a variety of scenarios:  
(i) $\nu_\mu - \nu_\tau$ (or $\nu_\mu - \nu$ sterile) mixing, 
(ii) $\nu_\mu - \nu_e$ mixing, (iii) three neutrino mixing
(iv) massless $\nu$ mixing etc.  

The general features of the asymmetry plot are
easy to understand.  For $\nu_\mu - \nu_\tau ($ or $\nu_\mu - \nu_{st})$ 
case, $A_\mu$ increases with energy, and
$A_e$ remains 0; for $\nu_\mu-\nu_e$ mixing, 
$A_e$ and $A_\mu$ have opposite signs; the three neutrino
cases interpolate between the above two; for the massless case the
energy dependence is opposite and the asymmetries decrease as $E_\nu$ 
is increased; when both $\nu_\mu$ and $\nu_e$  mix with sterile $\nu's$,
both $A_\mu$ and $A_e$ are positive etc.  With enough statistics, it
should be relatively straightforward to determine which is the correct
one.  Preliminary indications point to $\nu_\mu - \nu_\tau$
as the culprit.  There is also another suggestion \cite{vissani} 
which can in principle
distinguish $\nu_\mu -\nu_\tau$ from $\nu_\mu -\nu_{st}$ mixing.  If one
considers the total neutral current event rate divided by the total
charged current event rate; the ratio is essentially the n.c. cross
section divided by the c.c. cross section.  With  $\nu_\mu -\nu_{st}$
oscillations the ratio remains unchanged since $\nu_{st}$ has neither
n.c. nor c.c. interactions and the numerator and denominator change
equally $(\nu_\mu -  \nu_e$ case is even simpler:  nothing changes);
however, in $\nu_\mu - \nu_\tau$ case the denominator decreases and the
ratio is expected to increase by $\left ( \frac{1+r}{P+ r} \right )
\approx 1.5$, (here $r = N^0_{\nu e}/ N^0_{\nu \mu} \approx 1/2$ and
$P = 1/2 = \nu_\mu$ survival probability).  Of course, it is
difficult to isolate neutral current events; but it is proposed
to select $\nu N \ra \nu \pi^0 N$ and $\nu N \ra \ell \pi ^\pm N$ events
and the Kamiokande data seem to favor $\nu_\mu - \nu_\tau$ over 
$\nu_\mu -\nu_{st}$ or $\nu_{\mu} - \nu_{e}$ \cite{vissani}.

The new data from SuperKamiokande seem\cite{totsuka} to rule out all
non-oscillation explanations, prefer a value for $\delta m^2$ near 
$5.10^{-3} \ eV^2$ with large mixing and also prefer $\nu_\mu -
\nu_\tau$ over $\nu_\mu - \nu_e$. 
$\nu_\mu - \nu_e$ mixing is also excluded as a result of
the new CHOOZ data\cite{chooz}.

If we scale $L$ and $E$ each by the same amount, say $\sim 100$, we
should
again see large effects.  Hence, upcoming thrugoing $\mu's$ which
correspond to $E \sim 100$ GeV on the average, with path lengths of 
$L \tilde{>}$ 2000 km should be depleted.  There are data from Kolar
Gold Fields, Baksan, Kamiokande, IMB, MACRO, SOUDAN and now SuperK.  It
is difficult to test the event rate for $\nu_\mu$ depletion since there
are no $\nu_e's$ to take flux ratios and the absolute flux predictions
have 30\% uncertainties.  However, there should be distortions of the
zenith angle distribution and there seems to be some evidence for this.

\section{Solar Neutrinos}

Since the work of von Weizsacker \cite{weizacker}, Bethe
 \cite{bethe}  and Critchfield in the 30's and
40's, we believe that the energy of the sun is generated by conversion
of hydrogen to helium.  The basic reaction is
\begin{equation}
4p \ra  {^4{He_2}} + 2 e^+ + 2 \nu_e
\end{equation}
with a release of 25 MeV.  Most of the energy goes into producing
photons which eventually emerge as sunlight and the two neutrinos share
about 2 MeV.  From the fact that the energy density in sunlightt at
earth's surface is $1400 Jm^{-2} sec^{-1}$, it is easy to estimate that
the neutrino flux at the earth is about $10^{11} cm^{-2} sec^{-1}$.
These neutrinos, unlike the light, come directly from the center of the
sun, and probe the interior in a unique way.  The actual energy spectrum
and flux of the neutrinos depend on the intermediate steps in the
reaction above as  show in the Table~\ref{reaction} from the book by
Bahcall \cite{bahcall}.   

The first and  pioneering solar neutrino detector is the one build by
Davis and his collaborators.  It is a tank containing 100 000 cubic feet
of $CCl_4$ in Homestate gold mine in South Dakota.  The aim is to look
for the reaction $\nu_e + ^{37}Cl \ra e^- + ^{37}Ar$.  The Argon-37
decays with a half-life of 37 days.

\begin{table}
\caption{The pp  chain in the Sun.  The average number of pp neutrinos
produced per termination in the Sun is 1.85.  For all other neutrino
sources, the average number of neutrinos produced per termination is
equal to (the termination percentage/100).}
\begin{center}
\begin{tabular}{llcc}  \hline
Reaction  &   Number  &    Termination
                                       &  $\nu_e$ energy  \\
               &     &      (\%)            & (MeV)  \\ \hline 
$p + p \ra ^2H + e^+ + \nu_e$     &  1a              & 100     &  $ \leq$
0.420 \\
\mbox{or}      &      &   &    \\
$p + e^- + p \ra ^2H + \nu_e$    &   1 b (pep)     &  0.4    &   1.442
\\
$^2H + e^- \ra ^3He + \gamma$    &    2             &  100   &   \\
$^3He + ^3He \ra \alpha + 2p$    &    3             &  85    &  \\
\mbox{or}      &    &      &  \\
$^3He + ^4He \ra ^7Be + \gamma$    &   4            &   15   &  \\
$^7Be + e^- \ra ^7Li + \nu_e$       &   5           &    15  &  (90\%) 0.861 \\
    &    &    					             &  (10\%) 0.383 \\
$^7Li + p \ra 2 \alpha$             &  6            &    15   &   \\
\mbox{or}   &    &   &   \\
$^7Be + p \ra ^8B + \gamma$  	   & 7              &   0.02    &  \\
$^8B \ra ^8Be^{*} + e^+ + \nu_e$       & 8             &   0.02    &  $< 15$
\\
$^8Be^{*} \ra 2 \alpha$                &  9            &   0.02   &   \\
\mbox{or}   &  &   &  \\
$^3He + p \ra ^4He + e^+  + \nu_e$  &   10 (hep)    &   0.00002    &
$\leq$ 18.77 \\ \hline
\end{tabular}
\end{center}
\label{reaction}
\end{table}        

The tank is flushed every month with helium which flushes out the Argon
and then one looks for the radioactivity of $^{37}Ar.$  The number of
Argon atoms is extremely small, the total number detected in over twenty
years of running (1970-1997) is of the order of a few hundred.  The
average counting rate is almost 1/4th of the expected rate in the
standard solar model (SSM).

Another detector is the Kamiokande in a zinc mine in Kamioka, Japan.  It
consists of about 1000 tons of water surrounded by phototubes to detect
Cerenkov light emitted by charged particles.  The reaction being studied
is $\nu_e + e \ra \nu_e + e$ where the final electron emits Cerenkov
light and should be in the same direction as the initial $\nu_e$ and
hence point away from there sun.  Due to high background at low
energies, only electrons above 7.5 MeV can be detected.  In data
collected over 9 years (1987-1996), the observed rate is about 40\% of
the rate expected in SSM.  Since 1990, two Gallium detectors (Gallex at
Gran-Sasso and SAGE in Russia), have been taking data as well. They are
sensitive to low energy p p neutrinos via the low threshold reaction
$\nu_e +^{71}Ga \ra e^- + ^{71}Ge$.  The method is chemical and similar to
the Davis experiment.  The $^{71}Ge$ decays back to $^{71}Ga$ by
e-capture with a half-life of 11 days and the $^{71}Ga$ is extracted
chemically.  Gallex employs 30 tons of $GaCl_3$ solution whereas SAGE
uses 60 tons of metallic Gallium.  The current observed rate is about
50\% of the SSM expectation.

The data from four solar neutrino detectors (Homestake, Kamiokande, SAGE
and Gallex) have been discussed extensively \cite{homestake}.
The SuperK data are consistent with those from Kamiokande but increase
the statistics by an order of magnitude in one year
\cite{nakahata,totsuka}.  
To analyze these
data one makes the following assumptions:  (i) the sun is powered mainly by
the pp cycle, (ii) the sun is in a steady state, (iii) neutrino masses
are zero and (iv) the $\beta-$decay
spectra have the standard Fermi shapes.  Then it is relatively 
straightforward to show using these data with the solar luminosity that the
neutrinos from $^7Be$ are absent or at least two experiments are wrong \cite{hata}.
$^7Be$ is necessary to produce 
$^8B$ and the decay of $^8B$ has been observed; and the rate for $^7Be +
e^- \ra \nu + Li$ is orders of magnitude greater than $^7Be + \gamma \ra
^8B + p$ and hence it is almost impossible to find a ``conventional''
explanation for this lack of $^7Be$ neutrinos.  The simplest explanation
is neutrino oscillations.

Assuming that neutrino oscillations are responsible for the solar
neutrino anomaly; there are several distinct possibilities.  There are
several different regions in $\delta m^2 - sin^2 2 \theta$ plane that are
viable:  (i) ``Just-so'' with $\delta m^2 \sim 10^{-10} eV^2$ and
$\sin^2 2 \theta \sim 1$ \cite{glashow}, (ii) MSW small angle with 
$\delta m^2 \sim 10^{-5} eV^2$ and $\sin^2 2 \theta \sim 10^{-2}$ and
(iii) MSW large angle with $\delta m^2 \sim 10^{-7} eV^2$ (or  
$\delta m^2 \sim 10^{-5} eV^2)$
and $\sin^2  2 \theta \sim 1 $.  The ``just-so'' is characterized by
strong distortion of $^8B$ spectrum and large real-time variation of
flux, especially for the $^7Be$ line; MSW small angle also predicts
distortion of the $^8B$ spectrum and a very small $^7Be \nu$ flux and MSW
large angle predicts day-night variations.  
These predictions
(especially spectrum distortion) will be tested in the SuperK as well as SNO
detectors.  In particular SNO, in addition to the spectrum, will be able
to measure $NC/CC$ ratio thus acting as a flux monitor and reducing the
dependence on solar models.

The only way to directly confirm the absence of $^7Be$ neutrinos is by
trying to detect them with a detector with a threshold low enough in
energy.  One such detector under construction is Borexino, which I
describe below \cite{arpesella}. 

Borexino is a liquid scintilator detector with a fiducial volume of
300T; with energy threshold for 0.25MeV, energy resolution of 45 KeV and
spatial resolution of $\sim 20 cm$ at 0.5 MeV.  The PMT pulse shape
can distinguish between $\alpha's$ and $\beta's$.  Time correlation
between adjacent events of upto 0.3 nsec is possible.  With these
features, it is possible to reduce backgrounds to a low enough level to
be able to extract a signal from $^7Be \ \nu_e's$ via $\nu - e$
scattering.  Radioactive impurities such as $^{238}U$, 
 $^{232}Th$ and $^{14}C$ have
to be lower than $10^{-15}, 10^{-16} g/g$ and $10^{-18} (^{14}C/^{12}C)$
respectively.  In the test tank CTF (Counting Test Facility) containing
6T of LS, data were taken in 1995-96 and these reductions of background
were achieved.  As of last summer, funds for the construction of full
Borexino have been approved in Italy (INFN), Germany (DFG) and the
U.S. (NSF); and construction should begin soon.  The Borexino
collaboration includes institutions from Italy, Germany, Hungary, Russia
and the U.S..

	With a FV of 300T, the events rate from $^7Be \ \nu's$ is about 50
per day with SSM, and if $\nu_e's$ convert completely to $\nu_\alpha
(\alpha=\mu/\tau)$ then the rate is reduced by a factor 
$\sigma_{\nu \mu e}/\sigma_{\nu ee} \sim 0.2$ to about 10 per day, which is still
detectable.  Since the events in a liquid scintilator have no
directionality, one has to rely on the time variation due to the $1/r^2$
effect to verify the solar origin of the events.  If the solution of the
solar neutrinos is due to ``just so'' oscillations with 
$\delta m^2 \sim 10^{-10} eV^2$, then the event rate from $^7Be \ \nu's$ shows dramatic
variations with periods of months.

Borexino has excellent capability to detect low energy
$\bar{\nu}_e's$ by the Reines-Cowan technique:
$\bar{\nu}_e + p \ra e^+ + n, n + p \ra d + \gamma$ with 0.2
msec separating the $e^+$ and $\gamma$.  This leads to possible detection
of terrestial and solar $\bar{\nu}_e's$.  The terrestial $\bar{\nu}_e's$
can come from nearby reactors and from $^{238}U$ and
$^{232}Th$ underground.  The Geo-thermal  $\bar{\nu}_e's$ have a
different spectrum and are relatively easy to distinguish above reactor
backgrounds.  Thus one can begin to distinguish amongst various
geophysical models for the $U/Th$ distribution in the crust and mantle.
Solar  $\bar{\nu}_e's$ can arise via conversion of $\nu_e$ to
$\bar{\nu}_\mu$ inside the sun when $\nu_e$ passes thru a magnetic field
region in the sun (for a Majorana magnetic moment) and then
$\bar{\nu}_\mu \ra \bar{\nu}_e$ by the large mixing enroute to the earth \cite{raghavan}.

\begin{table}
\caption{Summary of current data on solar neutrinos}
\begin{center}
\begin{tabular}{lcc}  \hline
Expt.           &   $E_{th}$      &  Rate/SSM   \\ \hline
Homestake       &   0.8 MeV       &   $0.28 \pm 0.03$   \\  
Kamiokande    &    7.5 MeV       &   $0.42 \pm 0.06$   \\  
Gallex      &   0.2 MeV       &   $0.52 \pm 0.12$   \\
SAGE        &   0.2 MeV       &   $0.53 \pm 0.17$   \\   \hline
\end{tabular}
\end{center}
\end{table}

Among the detectors under construction is SNO \cite{mcdonald} (Solar Neutrino
Observatory) at Sudbury, Canada.  This is a Kiloton $D_2O$ Cerenkov
detector sensitive to the reactions.

\begin{equation}
{\nu_{e}} {^2_{1}} D \ra e^- pp, \ \nu_{\alpha} {^2_{1}} D \ra \nu p n  \
\mbox{and} \ \nu e \ra \nu e
\end{equation}
i.e. CC (charged current), NC (neutral current) and $\nu-e$ scattering,
respectively with energy threshold for electron energies about 5 MeV.
SNO can detect spectrum distortion quite clearly and can alo confirm
depletion of $\nu_e's$ by comparing NC to CC rates.  SNO is expected to
begin taking data within a year.  

LSND and 3 Neutrino Mixing:  We have not discussed the LSND experiment
in detail.  The LSND detector at Los Alamos used the neutrinos from $\pi
\ra \mu \nu_\mu, \mu \ra e \nu_e \nu_\mu$ to look for 
$\bar{\nu}_\mu \ra \bar{\nu}_e$ (and $\nu_\mu \ra \nu_e)$ conversion.
They reported a positive result \cite{lsnd} with $\delta m^2 \sim (0.5$
to $2) eV^2$ and $\sin^2 2 \ \theta \sim (1-2) 10^{-3}$.  The KARMEN
detector \cite{karmen} will be able to confirm this result within a few
months.  If these results hold up, one needs four neutrino states to
account for all the neutrino anomalies and hence at least one light
sterile state.

\section{Supernova Neutrinos}

In February 23, 1987, a supernova explosion was seen in the Large
Magellanic Cloud.  Neutrino signals were observed in the Kamiokande and
the IMB detectors.  Before discussing the observed signal, let me
briefly recapitulate what would be expected on general grounds.

The general picture of a type-II supernova goes like this \cite{raffelt}.  A red giant
star of mass greater than 10 solar masses reaches a stage when the core
implodes and neutronization occurs i.e., $e'$s and $p$'s combine to form
$n$'s and $\nu_e$'s: $e^- + p \ra n + \nu_e$.  Density can increase from
$10^{11}$ to $10^{14} g/cm^3$.  The energy released is about 1\%  of
the rest energy released and the process takes place in about $10^{-3}$
sec.  Subsequently the thermal neutrinos and antineutrinos (of all
kinds) are emitted via $e^+ e^- \ra \nu_i \ \bar{\nu}_i$.  These have
Fermi-Dirac energy distribution with temperatures of about 5 MeV for
$\nu_e, \bar{\nu}_e$ and 10 MeV for $\nu_\mu, \nu_\tau$.  The number of
$\nu$ and $\bar{\nu}$ are about equal, but the number of $\nu_\mu$ and
$\nu_\tau$ are about 1/2 of the number of $\nu_e$.  The time interval
over which the neutrino emission lasts is expected to be about 10
seconds or so.  The total energy emitted in neutrinos is about 10\% of
rest energy which corresponds to (2 to 4) $1 0^{53}$ ergs.

In water Cerenkov detectors such as IMB and Kamiokande, the reactions
possible are (i) $\bar{\nu}_e p \ra ne^+$ and (ii) $\nu_e \ra \nu_e$ and
most events would be from $\bar{\nu}_e p$ reaction since in $H_2 O$, the
event rate ratio for $(\bar{\nu}_ep)/(\nu_e)$ goes as $E_\nu/MeV$ and at
10 MeV (which is the expected mean energy of the neutrinos) is just 10.
The $\bar{\nu}_e p$ events should show no directional preference since
the cross-section is nearly isotropic whereas the $\nu_e$ events should
be forward peaked.

The neutrino events seen by Kamiokande and IMB are in remarkable
agreement with these general expectations \cite{hirata}.  The total number of events
seen over about 10 second interval was 19, the energies ranged between 7
and 30 MeV with a mean of about 15 MeV.  The total energy in
$\bar{\nu}$'s was about $3 \times 10^{52}$ ergs which translates into
$3 \times 10^{53}$ ergs in total emitted energy in neutrinos.  The
angular distribution is nearly flat as expected with a slight forward
preference and one event is perhaps better interpreted as a $\nu_e$
event.  A fit to Fermi (or Maxwell) distribution suggests a temperature
for $\nu_e$ in the range 3.5 to 5 MeV.  There was also detection of a few
events \cite{alexeyev} by the LSD detector in Mont Bianc and by a similar detector in
Baksan, but there are questions about how to interpret these events and
we shall ignore them.  I will now list the neutrino properties that can
be constrained tightly and uniquely by these few events.

(1) $\nu_e$ lifetime:  Since the expected number of $\bar{\nu}_e$'s arrived form LMC, they
lived at least as long as the flight-time.  The flight path is about 52
kiloparsec which corresponds to a
flight-time (at speed of light) of $5 \times 10^{12}$ sec.  Hence the
laboratory life-time for 10 MeV $\nu_e$'s is greater than $5 \times
10^{12}$ sec.  The limit in the rest-frame is $(m_\nu/10 \ MeV) 5
\times 10^{12}$ sec.  If $\nu_e$ is not a mass eigenstate but there is a
mixing, then at least one (the lowest) eigenstate should have (a) large
component of $\nu_e$ and (b) live longer than $5 \times 10^{12}$ sec.

(2) $\nu_e$ photonic decay:  In observations by a satellite during 
the 10 sec period of the SN
1987A neutrino burst, no gamma rays of energies in the range (1 to 10
MeV) were seen  \cite{oberauer}.  Since the number density of $\nu_e$'s due to SN1987a
was about $10^{10}$ per $cm^2$, one can plate a limit of $10^{22}$ sec
on $\tau_{\nu e}/B.R$, where B.R is the {\it branching ratio} of $\nu_e$
to decay into photons.  This limit is valid for $m_\nu < 200$ eV even
for $\nu_\mu$ and $\nu_\tau$

(3) $\nu_e$ mass:  From the data we know that largest spread in arrival
times is $\Delta t \sim 10$ sec and the largest energy spread is between
10 and 30 MeV.  If two neutrino arrival time difference is $\Delta
t_{21}$, this can come from two sources: for a non zero mass of $\nu_e$,
the energy difference gives velocity difference giving an arrival time
difference and there may be a departure time difference $\delta t_{21}$;
i.e.:
\begin{equation}
\Delta t_{21} = \frac{R_{LMC}}{c} \frac{1}{2}
  (m_\nu c^2)^2 \ (E^{-2}_2 - E_1^{-2}) + \delta t_{21}
\end{equation}

If we assume that $\Delta t_{21} >> \delta t_{21}$, then
\begin{equation}
m_\nu c^2 \leq \left (
\frac{2 c \Delta t_{21}}{R_{LMC}}
\frac{E^2_2 E^2_1}{E^2_2 - E^2_1} \right )^{1/2}
\end{equation}
which for $E_1, E_2 \sim 10,$ 30 MeV and $\Delta t_{21} \sim 10$ sec,
gives a limit of about 20 eV.  More sophisticated analysis can not do
much better.  

(4) Mass of $\nu_\mu$ and $\nu_\tau$:  If $\nu_\mu$ and $\nu_\tau$ have
Dirac masses, then the neutral current scattering on nuclei changes
chirality and can create non-interacting sterile right handed $\nu$'s at
a rate proportional to $m^2_\nu$.  If $m^2_\nu$ is too large, this
happens in a short time scale and $\nu_R$'s would escape carrying away
too much energy.  From such considerations, one can put an upper bound
on $\nu_\mu, \nu_\tau$ masses of order 100 KeV or so \cite{gandhi}.  This is much
better than the laboratory bounds mentioned earlier.

(5) Electric charge of $\nu_e$:  If $\nu_e's$ had an electric charge of
$Q$, they would be deflected in the galactic magnetic fields according to
their energies leading to dispersion in arrival times.  The fractional
deviation from a straight line path $\Delta s/s$, can be shown to be 1/24
$(s^2/R^2)$ where $R$ is the radius of the path and is given by E/c QB
where $E$ is the neutrino energy, and $B$ is the magnetic field.  The
dispersion in arrival times is then $\Delta t \sim t \Delta (\Delta
s/s)$, where $t$ is the flight time of $5.10^{12}$ sec. For an observed
dispersion time of 10 sec, one can find a limit on $Q$, the neutrino
charge given by \cite{barbiellini}
\begin{equation}
Q \leq (\Delta t/t) \surd 24 [\langle E \rangle / 0.3 s \ B ] \surd
 \{E/2 \Delta E\}
\end{equation}
where $\langle E \rangle$ in GeV is about $15 \times 10^{-3}, s$ is $1.5
\times 10^{21}$ m, $B$ is $10^{-10}$ in tesla, $E/2 \Delta E$ is about 1
and the bound on $Q$ is about $10^{-14} \mid e \mid$.  The laboratory
bounds on $Q$ are stronger but  depend on charge conservation and charge
additivity, whereas this measures charge dynamically.

(6) Neutrino speed:  According to the earliest optical observation, the
time difference between the arrival of neutrino and photons from
SN1987A is less than a few hours or $2 \times 10^{4}$ sec.  Hence
\begin{equation}
\Delta t = t_\gamma - t\nu \ \tilde{<} 2 \times 10^4 sec
\end{equation}
So if neutrino speed is $v_\nu$, then
\begin{equation}
\mid 1 - \upsilon_\nu /c \mid < \Delta t /t = 5 \times 10^{-9}
\end{equation}
and hence $\upsilon_\nu = c$ to within a few parts in a billion \cite{stodolsky}.

(7) Neutrino Flavor:  If the number of light $(\leq 1$ MeV) neutrino
flavors were $N$, then the luminosity in $\tilde{\nu}_e$'s would be
reduced by 3/N from expected.  But the observed luminosity was just what
was expected, hence if we allow for a factor of 2 uncertainty, then $N$
should be less than 6. This limit has been
superseded by the LEP results on $Z^0$ which put $N=3$ very
accurately.

(8) Equivalence Principle for $\nu_e$:  If $\nu_e$ feels the
gravitational interaction due to our galaxy, then their time delay
should be.
\begin{equation}
\Delta t_\nu = M(1 + \delta_\nu) 1n \ (2R/b)
\end{equation} 
where $M$ is the mass of Milky Way and $b$ is the distance of solar
system from the center; $\delta_\nu$ should be one in general
relativity.  The time delay for photon is given by a similar formula
with $\delta_\gamma$ instead of $\delta_\nu. \delta_\gamma$ is known
that
\begin{equation}
\Delta t = \Delta t_\nu - \Delta t \gamma \tilde{<} 2 \times 10^4 sec
\end{equation}
and $M$  (in appropriate units!) $\sim 3 \times 10^5 sec, R/b \sim 4,
\Delta t_\nu \sim 3$ months, and hence
\begin{equation}
\frac{1}{2} (\delta_\nu - \delta_\gamma) = (\Delta t_\nu/ \Delta
t_\gamma)
\leq 2 \times 10^4 \mbox{sec/5 months} \sim 10^{-3}
\end{equation}
Hence $\delta_\nu = \delta_\gamma$ to within 1 part in 1000 \cite{longo}.

(8) Equivalence Principle for $\nu_e$ and $\bar{\nu}_e$:  In addition, 
if one interprets the one events as being due to
$\nu_e$ e scattering, then since the time interval is about ~ 1 sec, one can
test particle-antiparticle equivalence between $\nu_e$ and
$\tilde{\nu}_e$ \cite{pakvasa}
\begin{equation}
\frac{1}{2} (\delta_\nu -\delta_{\tilde{\nu}}) \tilde{<} 10^{-6}
\end{equation}

(9) New Forces on $\nu_e's$:  The above result can 
also be used to place bounds on new forces
(long range) coupling to neutrinos \cite{pakvasa}.  The potential energy between
neutrino and the galactic matter is given by
\begin{equation}
E(r) = - \left (
\frac{2G E_\nu m_2}{r} \pm
\frac{q_1 q_2}{r} +
\frac{m_\nu S_1 S_2}{r}
\right )
\end{equation}
where $m_2$ is galactic mass, $m_y$ the neutrino mass; $q_i$ are the
neutrino and galactic charges for a vector force-field, $S_i$ are for a
scalar force-field and the upper (lower) sign refers to neutrinos
(anti-neutrinos).  Then a bound on the vector force can be  obtained
from
\begin{eqnarray}
\delta t_\nu - \delta t_{\bar{\nu}} & = &
(m_\nu/E_\nu)^2
(2q_\nu Q \ E_\nu) [-R/\surd (R^2 + p^2) ] \nonumber \\
 &  & - 1n \{(R + \sqrt{R^2 + b^2})/b\} ] \leq 1 sec
\end{eqnarray}
For example, if vector field couples to a combination of $B$ and $L$
i.e. $G= g (\alpha L + \beta B) \surd C_5$, then one finds that $g^2
\alpha (\alpha + 1.2 \beta) < 2 \times 10^{-3} \ GeV^2$ for $m_\nu \sim
15$ eV.  If coupling to neutrino and to other matter is different, then
\begin{equation}
g^2 \rho (\alpha \delta + \beta) < 2 \times 10^{-3} GeV^2
\end{equation}
and for the dimensionless coupling $f_\nu$ defined by
\begin{equation}
g=f_\nu \surd G, f^2_\nu \rho ( \alpha \delta + \beta) < 2 \times
10^{-35}
\end{equation}
For the scalar force, which contributes only to the time delay between
$\nu$'s and $\gamma$'s, one finds $f^2_s, \rho ( \alpha \delta + \beta)
< 10^{-34}$.  If neutrinos from the dark matter in the galaxy, the bounds
are stronger:  $f^2_\nu \rho^2 < 3 \times 10^{-43}$ and
$f^2_s \rho^2 < 10{-32}$.

(10) Secret Interactions of $\nu_e's$:  Any new interactions of
$\nu_e$'s with majorons or self interactions cannot be too strong,
otherwise the number arriving would have been affected.  This is because
$\nu$'s can scatter off $\nu$'s in the 3 K neutrino background
radiation.  The limits on coupling constants obtained this way are
typically of the order of $10^{-3}$ \cite{kolb}.

(11) Mixing Angles of Neutrino:  Generally no information on the mixing
angles or mass difference of $\nu$'s can be obtained from the supernova
data.   But if one event is assumed to be $\nu_e e$ scattering, then the
matter effects in the supernova place strong constraints on $\delta m^2,
\sin^2 (2 \theta)$ that are allowed , in order that $\nu_e$ flux be
not depleted.  When superimposed on the MSW solution for the solar
neutrinos, it disfavors the large angle solution \cite{arafune}.

(12) Neutrino Magnetic Moment:  Before the core cooled to $T \sim$ 5 MeV,
it had a $T \sim $ 50-100 MeV.  If $\nu$ had a magnetic moment, when
$e^+ e^- \ra \nu_R \bar{\nu}_R$ by the magnetic interaction and the
$\nu_R$ would escape (whereas $\nu_L$ of this energy is trapped by its
small mean free path).  This would create two problems:  one that no
neutrinos of this energy were observed $(\nu_R \ra \nu_L$ in the galactic
magnetic fields) and the other is that so much energy would be lost that
there would be little left in the low energy $\nu_L$'s.  If  this
argument were valid \cite{barbieri}, one would obtain bounds on $\nu$ magnetic dipole
moment of $10^{-12}\mu_{\mbox{Bohr}}$.  However, if the neutrino is a
Majorana particle or if the $\nu_R$'s have new interactions, this
argument breaks down since $\nu_R$ would then be trapped.  So strictly
speaking, there is no limit on the neutrino magnetic moment from SN
1987a.

For a paltry 19 events, this is a tremendous amount of information on
neutrinos.  
We hope that the next observed supernova would be inside Milky Way   
at a distance under 10 kiloparsec.  Then the neutrino events seen would
be in the order of several hundred.  Can this happen in the next twenty
years or so?  Supernova watch is continuing.

\section{Early Universe}

The cosmic microwave background radiation should be accompanied by
neutrinos which decoupled at very early times \cite{kolb2}.  The present temperature
of neutrino $T_\nu$ is related to the photon temperature $T_\gamma$ by 
\begin{equation}
T_\nu = (4/11)^{1/3} \ T_\gamma
\end{equation}
which gives $T_\nu = 1.9$ K for $T_\gamma$ of 2.7 K.  The physical
reason for the difference by the factor $(4/11)^{1/3}$ is the raising of
photon temperature after decoupling due to $e^+ e^-$  annihilation which
dumps energy in photons.  Assuming a Fermi-Dirac distribution a
temperature of 1.9 K yields for the neutrino density
\begin{equation}
n_\nu = 115 \ per \ cc
\end{equation}
for each flavor.  Hence the energy density in neutrinos is
\begin{equation}
\rho_\nu = \sum_{i} 115 (m_{\nu_{i}} c^2) \ \mbox{per cc}
\end{equation}
For this to be less than the critical density $\rho_c = 10^4 h^2 \
eV/cc$, the sum of all neutrino masses must satisfy (for h $\sim 0.4$ to 1).
\begin{equation}
\sum_{i} m_{\nu_{i}} < 100 \ eV
\end{equation}
This is the well-known result \cite{cowsik} 
due to Cowsik, McLeland, and Zeldovich.

These results raise two questions.  One is the fact that if neutrino
masses add up to 10-30 eV  (or if one neutrino, say $\nu_\tau$ has such
a mass) then neutrinos can provide the bulk of the energy density of the
universe and a sizable fraction of the dark matter (30 to 100\%) in the
form of hot dark matter.  How can this be tested experimentally?  The
other is, whether these cosmic background neutrinos themselves be
detected experimentally?

For the detection of relic $\nu$'s many suggestions have been made over
the years.  None of them is in the danger of being implemented in the
near future.  Some early proposal were based using coherent surface
effects and detect a net force, on a large area due to the earth's
motion in space.  This turned out to be too small for detection after it
was shown that the effect is proportional to $G^2_F$ rather than $G_F$.
Another elegant idea \cite{stodolsky2} is to use a possible $\nu - \bar{\nu}$ asymmetry to
create an effective parity violating force on polarised electrons and
look for the spin rotation with propagation.  At very low temperatures
with all external magnetic fields quenched, this may become feasible
some day.  Another suggestion \cite{weiler} is to use very distant sources of very high energy 
$\nu$'s and look for Z-absorption dips in their spectrum due to the
process $\nu + \nu_{CBR} \ra Z^0$.  Perhaps the most promising and
practical proposal is the one due to Zeldovich et al \cite{shvartzman}.  They propose
using the volume effect by employing a loosely filled container (about
half-filled) with spheres of size $\alpha \sim 0 (\lambda)$ when $\lambda$ is the
wavelength of the CBR $\nu$'s and keeping the interstitial distance $d$
less than $\lambda$.  The acceleration experienced by the container is
\begin{equation}
a \sim 10^{-22} \ (K_L/A)^2 \ cm/sec^2
\end{equation}
for neutrino mass in the eV range. Here $K_L = (A -Z)$   for $\nu_\mu$
and $(3 Z - A)$ for $\nu_e$.  Can such objects be constructed and such
small accelerations be measured?

If one neutrino, say $\nu_\tau$ has a mass in the range 5 to 30 eV it
can provide 30 to 100\% of energy density needed to guarantee a $\Omega$
of 0(1).  If such a $\nu_\tau$ mixes with $\nu_\mu$ by an amount
$>10^{-4}$  ongoing and future experiments at CERN and Fermilab such as
CHORUS, NOMAD and COSMOS should be able to confirm that.  These are
appearance experiments in which if $\nu_\mu \ra \nu_\tau \ra \nu_\tau +
N \ra \tau + \ x$ takes place, they can be detected.  The sensitivity is
to probe $\delta m^2 \sim 1$ to $1000 \ eV^2$ and $\sin^2 2 \theta$ up
to $10^{-4}$.

\section*{Conclusion and Summary}

We expect great progress in this field in the next 4 or 5 years and hope
for eventual unambiguous evidence for physics beyond the standard model
from neutrino properties.

The neutrinoless double beta decay limits should be pushed to at least
as low as 0.1 eV.  The new solar neutrino experiments with rates of
several thousand events per year should confirm (or deny) the anomaly
and measure $\delta m^2$ and mixing angles.  Long baseline experiments
(as well as Superkamiokande) should settle the question of $\delta
m^2$ near $10^{-2} eV^2$ with large mixing for either $\nu_\mu - \nu_e$
or $\nu_\mu - \nu_\tau$.  Short baseline experiments at CERN and
Fermilab should check $\nu_\mu -\nu_\tau$ oscillations with large
$\delta m^2$  and $\sin^2 \ 2 \theta$ upto $10^{-3} - 10^{-4}$ and thus
indirectly the identity of Hot Dark Matter.  If we are fortunate we may
have a Galactic Supernova and we may be about to witness the early days
of an emerging new field:  high energy neutrino astronomy.  These are
exciting times.

\section*{Acknowledgment}

I thank Alexei Smirnov, Antonio Masiero and Goran Senjanovic for the
invitation to lecture in Trieste, for a most stimulating atmosphere in
the school and outstanding hospitality.  This work was supported in part by
U.S.D.O.E. under Grant \#DE-FG-03-94ER40833.


\section*{References}
\begin{thebibliography}{99}
\bibitem{wit}B. de Wit and D.Z. Freedman, Phys. Rev. Lett. 95, 827
(1975).

\bibitem{yanagida} T. Yanagida, {\it Proc. of Workshop on Unified Theory and
Baryon Number of the Universe}, ed. by O. Sawada and A. Sugamoto, KEK
(1979); M. GellMann, P. Ramond and R. Slanksky, Supergravity, ed. by
P. von Nienvenhuizen and D.Z. Freedman, N. Holland (1979).

\bibitem{gelmini} G. B. Gelmini and M. Roncadelli, {\it Phys. Lett.} 99B, 411 (1981).

\bibitem{chicasige}Y. Chicasige, R.N. Mohapatra and R. Peccei,
{\it Phys. Lett.} 98B, 265 (1981).

\bibitem{lobashov}Talks by I. Nikolic and by 
V.M. Lobashov, {\it Int. Workhop on Weak Interactions and
Neutrinos}, Capri, June 1997 (Elsevier), to be published; K. Assamagan et
al. {\it Phys. Rev.} D53, 6065 (1996). 

\bibitem{kayser} B. Kayser, {\it Phys. Rev.} D24, 110
(1981).

\bibitem{kim} C. W. Kim and A. Pevsner, {\it Neutrinos in Physics and
Astrophysics}, Harwood (1994).

\bibitem{cabibbo} N. Cabibbo, {\it Phys. Lett.} 72B, 333 (1978); L. Wolfenstein,
Phys. Rev. D18, 958 (1978).

\bibitem{lee}B. W. Lee, S. Pakvasa and H. Sugawara, {\it Phys. Rev. Lett.} 38,
937 (1977); S. B. Treiman, F. Wilczek and A. Zee, Phys. Rev. D16, 152
(1977).

\bibitem{nakagawa} M. Nakagawa et al, {\it Proc. Theor. Physics} 30, 258
(1963).

\bibitem{gasperini}M. Gasperini, {\it Phys. Rev.} D38, 2635 (1988); A. Halprin
and C. N. Leung, {\it Phys. Rev. Lett.} D67, 1833 (1991).

\bibitem{coleman}S.  Coleman and S. L. Glashow, {\it Phys. Lett.}  B405, 249
(1997).

\bibitem{wolfenstein} L. Wolfenstein, {\it Phys. Rev.} D17, 2369 (1978).

\bibitem{mikheyev} S. P. Mikheyev and A. Y. Smirnov, {\it Nuov. Cim.} C9, 17
(1986).

\bibitem{berezhiani} V. Berezhiani and M. I. Vysotsky, {\it Phys. Lett.} 199B,
281 (1987).

\bibitem{fujikawa}K. Fujikawa and R. Shrock, {\it Phys. Rev. Lett.} 45, 963 (1980).

\bibitem{haxton} W.C. Haxton and G. Stephenson, {\it Progr. Part. Nucl. Phys.} 12, 408 (1989).

\bibitem{nakahata} M. Nakahata, {\it Proceedings of PPPP Symposium}, Seoul,
Nov. 1997, World Scientific (to be published).

\bibitem{totsuka}Y. Totsuka, {\it Proceedings of the Lepton Photon
Symposium}, Hamburg, July 1997 (to be published).

\bibitem{kasuga}S. Kasuga et al., {\it Phys. Lett.} B374, 238 (1996).

\bibitem{volkova}L. Volkova, {\it Phys. Lett.} B316, 178 (1993).

\bibitem{ryazhkaya}O. G. Ryazhkaya, {\it JFTP Lett.} 61, 237 (1995).

\bibitem{fukuda}Y. Fukuda et al., {\it Phys. Lett.} B388, 397 (1996).

\bibitem{bellotti}R. Bellotti et al., {\it Phys. Rev.} D53, 35 (1996).

\bibitem{learned}J. G. Learned, S. Pakvasa and T. J. Weiler,
{\it Phys. Lett.} B207, 79 (1988); V. Barger and K. Whisnant, 
{\it Phys. Lett.} 209B, 360 (1988); K. Hidaka et al., {\it Phys. Rev. Lett.}
61, 1537 (1988).

\bibitem{flanagan}J. Flanagan, J. G. Learned and S. Pakvasa, 
hep-ph/9709438, {\it Phys. Rev. D} in press.

\bibitem{vissani}F. Vissani and A. Smirnov, hep-ph/9710565.
 
\bibitem{chooz} The CHOOZ collaboration, M. Apollonio et al.,
hep-ex/9711002.

\bibitem{weizacker}C. F. Von Weizacker, {\it Zeit, fur Physik} 39, 663
(1938).

\bibitem{bethe}H. A. Bethe, Phys. Rev. 55, 434 (1939).

\bibitem{bahcall}J. N. Bahcall, ``{\it Neutrino Astrophysics,}''
Cambridge, (1989).

\bibitem{homestake} See talks by the Homeskate, Kamiokande, Sage and
Gallex Collaborations in:  {\it Proceedingss XVII International Conference on
Neutrino Physics and Astrophysics}, Helsinki Finland (13-19 June 1996),
eds. J. Maalampi and M. Roos (World Scientific, Singapore, 1997), to be
published.

\bibitem{hata} N. Hata and P. Langacker, Phys. Rev. D56, 6107 (1997);
J. N. Bahcall, hep-ph/9711358; V. Castellani et al. hep-ph/9606180;
S. Parke, {\it Phys. Rev. Lett.} 74, 839 (1995); K.V.L. Sarma,
hep-ph/9408277.

\bibitem{glashow}S.L. Glashow anddd L. M. Krauss, {\it Phys. Lett.} 190B, 199
(1987); V. Barger, R. J. N. Phillips and K. Whisnant, {\it
Phys. Rev.} D234, 528 (1981).

\bibitem{arpesella}C. Arpesella et al. {\it The Borexino Proposal}
Vol. 1 and 2 ed. G. Bellini and R. S. Raghavan (Univ. of Milan) 1991;
G. Alimonti et al. ({\it Nucl. Instruments \& Methods, in press}).

\bibitem{raghavan}R. S. Raghavan et al, {\it Phys. Rev. Lett.} 80, 635 (1998).

\bibitem{mcdonald}A. B. McDonald, {\it Proc. The Nineth Lake Louise Winter
Inst.}, ed. A. Astbury et al, World Scientific, 1994, p. 1.
\bibitem{lsnd}The LSND collaboration, C. Athanassapoulos et al, {\it
Phys. Rev.} C54, 2685 (1996) and nucl-ex/9709006.

\bibitem{karmen}The KARMEN Collaboration, K. Eitel et al. Proc. of the
32nd Rencontre de Moriond, Electroweak Interaction and Unified Theories,
Les Arcs, March 1997 (to be published), hep-ex/9706023.

\bibitem{raffelt}G. G. Raffeltt, {\it Stars as Laboratories for
Fundamental Physics}; University of Chicago Press, Chicago (1996).

\bibitem{hirata}K. Hirata et al. {\it Phys. Rev. Lett.} 58, 1490 (1987);
R. Bionta et al. ibid. 58, 1494 (1987).

\bibitem{alexeyev}E. N. Alexeyev et al. {\it JETP Lett.} 45, 589 (1987).

\bibitem{oberauer}L. Oberauer and F. von Feilitzsch, {\it Phys. Lett.}
B200, 580 (1988).

\bibitem{gandhi}R. Gandhi and A. Burrows, {\it Phys. Lett.} B246, 149
(1990).

\bibitem{barbiellini} R. Barbiellini and G. Cocconi, {\it Nature}, 329,
21 (1987).

\bibitem{stodolsky}L. Stodolsky, {\it Phys. Lett.} 201B, 353 (1988).

\bibitem{longo}M. Longo, {\it Phys. Rev. Lett.} 60, 173 (1988);
L. Krauss and S. Tremaine, ibid, 60, 176 (1988).

\bibitem{pakvasa}S. Pakvasa, W.A. Simmons and T.J. Weiler, {\it
Phys. Rev.} D39, 1761 (1989), J. Losecco, {\it Phys. Rev.} D38, 3313
(1988).

\bibitem{kolb} R. Kolb and M. Turner, {\it Phys. Rev.} D30, 2895 (1987).

\bibitem{arafune}J. Arafune et al, {\it Phys. Rev. Lett.} 59, 1564
(1987).

\bibitem{barbieri} R. Barbieri and R. N. Mohapatra, {\it
Phys. Rev. Lett.} 61, 27 (1988).

\bibitem{kolb2}R. Kolb and M. Turner, {\it The Early Universe},
Addision-Wesley California (1990).

\bibitem{cowsik} R. Cowsik and J. McLeland, Phys. Rev. Lett. 29, 669
(1972); S. Gerstein and Ya. B. Zeldovich, {\it JETP Lett.} 4, 174 (1966).

\bibitem{stodolsky2} L. Stodolsky, {\it Phys. Rev. Lett.} 34, 110 (1975).

\bibitem{weiler}T. J. Weiler, {\it Phys. Rev. Lett.} 49, 234 (1984).

\bibitem{shvartzman}V. F. Shvartzman et al. {\it JETP Lett.} 36, 224 (1983).  


\end{thebibliography}
\end{document}